\documentclass[12pt,preprint]{aastex}

\def\lsim{\hbox{ \rlap{\raise 0.425ex\hbox{$<$}}\lower 0.65ex\hbox{$\sim$} }}
\def\gsim{\hbox{ \rlap{\raise 0.425ex\hbox{$>$}}\lower 0.65ex\hbox{$\sim$} }}

\def\opeqn{\begin{equation}}
\def\cleqn{\end{equation}}

\def\dfrac{\displaystyle\frac}
\def\dint{\displaystyle\int}

\pdfoutput = 1

\begin{document}

\title{Perturbation of Gravitational Lensing}
\author{Sun Hong Rhie }
\affil{eplusminus@gmail.com}
\author{Clara S. Bennett}
\affil{Physics Department, Massachusetts Institute of Technology}


\begin{abstract}
A gravitational lens system can be perturbed by ``rogue systems" in 
angular proximities but at different distances. 
A point mass perturbed by another point mass can be considered as a
large separation approximation
of the double scattering two point mass (DSTP) lens. 
The resulting effective lens depends on whether the perturber is 
closer to or farther from the observer than the main lens system.
The caustic is smaller than that of the large separation binary lens
when the perturber is the first scatterer;  
the caustic is similar in size with the large separation binary lens 
when the perturber is the last scatterer. 
Modelling of a gravitational lensing
by a galaxy requires extra terms other than constanst shear 
for the perturbers at different redshifts. 
Double scattering two distributed mass (DSTD) lens
is considered. The perturbing galaxy behaves as a monopole 
-- or a point mass -- because the dipole moment of the elliptic mass
distribution is zero.   
\end{abstract}


\clearpage

\tableofcontents

\section{Introduction}

The Galactic bulge is being surveyed for gravitational 
microlensing in search of 
microlensing planets. The lensing systems are the standard single
scattering $n$-point mass lenses, where the bound system 
may consist of a single star,
a multiple star, or a planetary system with one or two host stars. 
The Galactic bulge microlensing probability is $\sim 10^{-6}$ and 
the probability for an unbound system to be aligned with the main
lensing system can be ignored because it is $\sim 10^{-12}$. 
The number of stars being surveyed is less than $10^9$.
The lensing probability is proportional to the Einstein ring
radius square, and the Einstein ring radius of the lensing toward
the Galactic Bulge is characteristically $\sim 1$ mas. Thus, if we
consider the probability of a ``rogue" system to be within $1$ as
from the main lensing system, the probability is $\sim 1$. 
In fact, ground-based microlensing events are known to be ``plagued" 
with blending of light, 
and some of them other than the main lens itself can also
be gravitationally relevant for the photon path. 
The perturbers can be dark as well.

\begin{figure}
\plotone{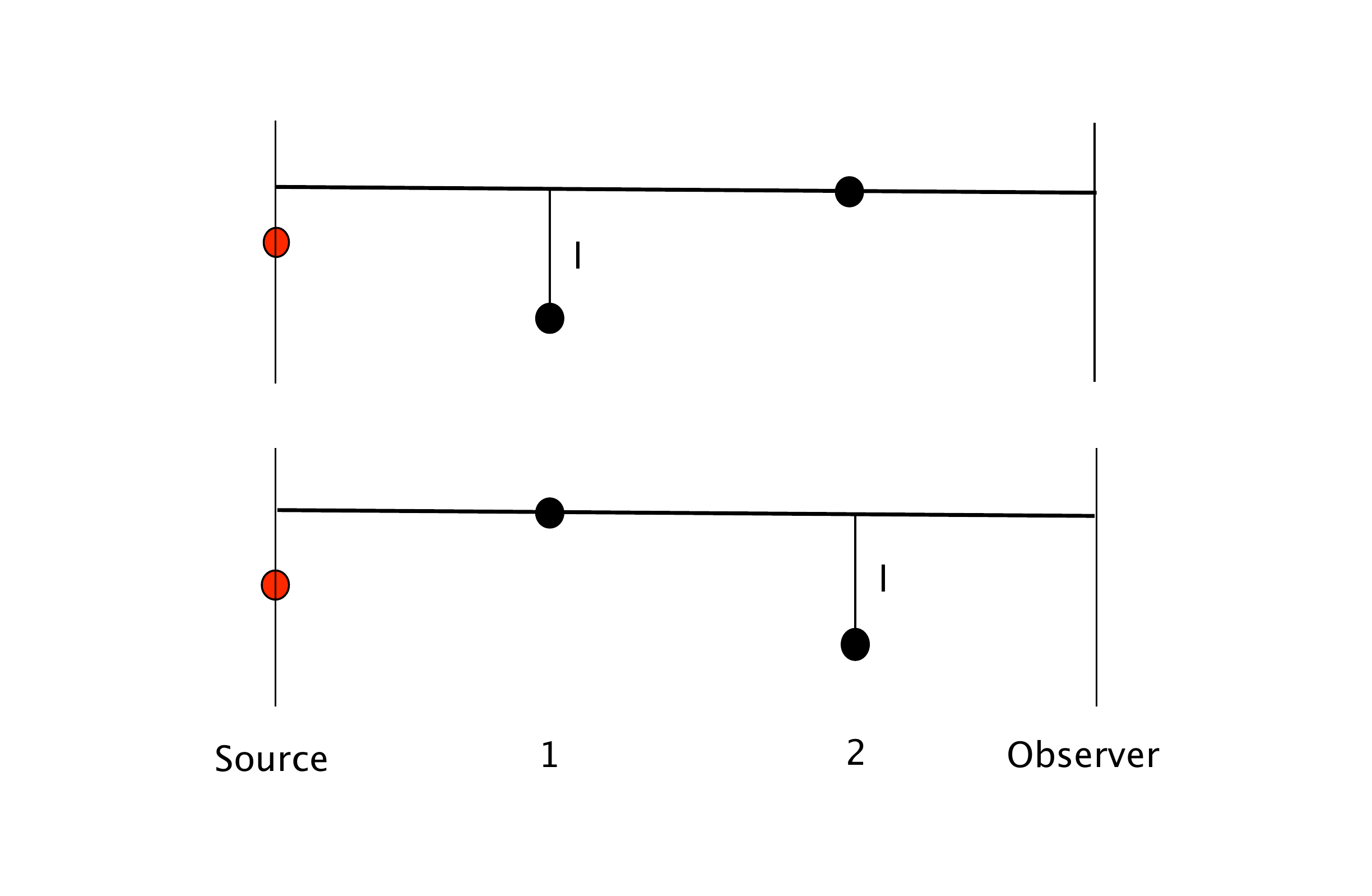}
\caption{Double scattering lensing is a time-sequential
process and there are two cases: 1) lens 1 is the perturber;
2) lens 2 is the perturber.
}
\label{fig_twocases}
\end{figure}

The probability for the ``rogue" system to be at the same distance
as the main lens system is small where the ``same distance" should
be understood in the context that 
the two lens systems are within the coherence scale 
in which the lensing can be considered a single scattering lensing. 
Within the coherence scale, the probability for the photon path 
to weave through the two lenses can be ignored \citep{RCB} 
(RB10 from here on).  Thus it is most reasonable to assume in general 
that the two gravitationally unbound microlens 
systems are at different distances, and they would be best considered 
as a double scattering lensing system. 
Here it is assumed that the two lens elements are widely 
separated in the sky 
based on the argument in the previous paragraph and
calculate the effects of the ``rogue" system  on the main lens.    
The double scattering lensing is a time-sequential process, and  
it matters whether the ``rogue" system is farther away than the
main lens from the observer or closer. The two cases are schematically 
shown in figure \ref{fig_twocases}. 
By wide separation it is implied that
the separation $\ell$ is much larger than the Einstein ring radius 
of the main lens.

We assume that the ``rogue" perturbing system is a single point mass. 
The main lens of interest will be a single star, a multiple star,
a planet system with one or two host stars, or a wide binary stars one
of which hosts planets. Here we consider the most common and simplest case 
of a signle star and study the wide separation approximation of the
double scattering two point mass (DSTP) lens.  Then the most important 
effect of the perturbation is to break the degeneracy of the point caustic
of the single lens to an extended caustic curve, and the size of the
caustic will be the indicator of the influence of the perturber. 
It will be shown that 
the caustic size depends on the distances of the lenses
and whether the perturber is in the front or in the back. 
When the perturber is the first scatterer,
the caustic is smaller than that of the binary lens, and it is 
similar in size when the main lens is perturbed by a ``rogue" system 
in front. 
A binary lens forms
when the two lenses have the same distance -- or within the coherence
length.   The binary lens
at large separation is made of a point mass and a constant shear
(and plus the source shift), and it is briefly discussed in the appendix.

A multiple-point mass lens perturbing a single point mass 
main lens is approximated by the same form of the approximate DSTP 
lens equation (with effective coefficients) and can be concluded to 
behave in the similar manner to the single point mass perturber.

In lensing by a galaxy, 
modelling is done customarily by assuming an elliptic
mass (sometimes replaced by an elliptic potential) 
and a constant shear. The galaxy lensing
is of order $1$ arcsecond, and there are often other galaxies in the 
angular vicinity of the main lens system.  The perturbers can be 
group memebers of the main lens or galaxies at different distances.
If there is a
perturbing galaxy at the same distance, its monopole  will add 
an external shear as is the case with the binary lens. However,
if the perturbing masses are at different distances, 
the perturbation should include deflection terms other than the 
constant external shear.
We consider the wide separation approximation of the 
double scattering two distributed mass (DSTD) lens.
The other terms depend on the double scattering parameter and vanish
when the perturbers are at the same distance as the main lens because 
the double scattering parameter vanishes. The perturbation by an elliptic
mass galaxy is approximated by the perturbation by a point mass (monopole) 
because the dipole moment of the elliptic mass distribution is zero.
The main lens galaxy, assumed to have an elliptic mass distribution,  
has a finite size caustic, and the effect of the perturber is to 
change its shape, size, and position. Even in the simpler case of the
main galaxy as the monopole-quadrupole lens requires numerical 
calculations. We leave the perturbations of the finite size caustics
for future work.

It should be necessary to point out that 
\citet{keeton} uses Taylor expansion and concludes that the effects  
of a perturbing mass on the galaxy lens is to add constant convergence 
and constant shear irrelevantly of whether the perturber 
is the first scatterer or the last. There may be a problem in the
expansion cutoff. In the region of interest around the critical curve
of the galaxy lens, the first term in the Taylor expansion is small
because the Jacobian determinant is zero or small and is likely to
be smaller than the second order term. The second order term is well
known for the square root behavior of the lensing near the critical 
curve or caustic crossing. It is not clear whether the
Taylor expansion can be used at all. We use power expansion around
the critical curve of the main lens which is the region of interest.

The DSTD lens equation is an obvious extension of the DSTP lens equation
in which the delta function integral for the 2-d
gravitational field of a point mass is 
generalized to the density function integral for the 2-d gravitational
field of the distributed mass.

The DSTP lens equation is known since 1986 (Blandford and Narayan)
and have been studied \citep{KA86, ES93, werner}. Here the derivation of
the DSTP lens equation studied in RB10 is briefed 
for clarity and convenience. Instead of using the formula for the time delay 
and Fermat principle, the well-known derivation of the single lens equation 
from an exact solution of the general relativity, the Schwarzschild metric,
with the assumption of the linear 
gravity and small angle approximation is used. 
The Schwarzschild metric is asymptotically flat and the scattering 
planes can be joined easily in the asymptotic regions.
The DSTP lens equation is obtained by joining two scattering planes 
with the freedom to rotate.

\section{The DSTP Lens Equation}

\begin{figure} 
\plotone{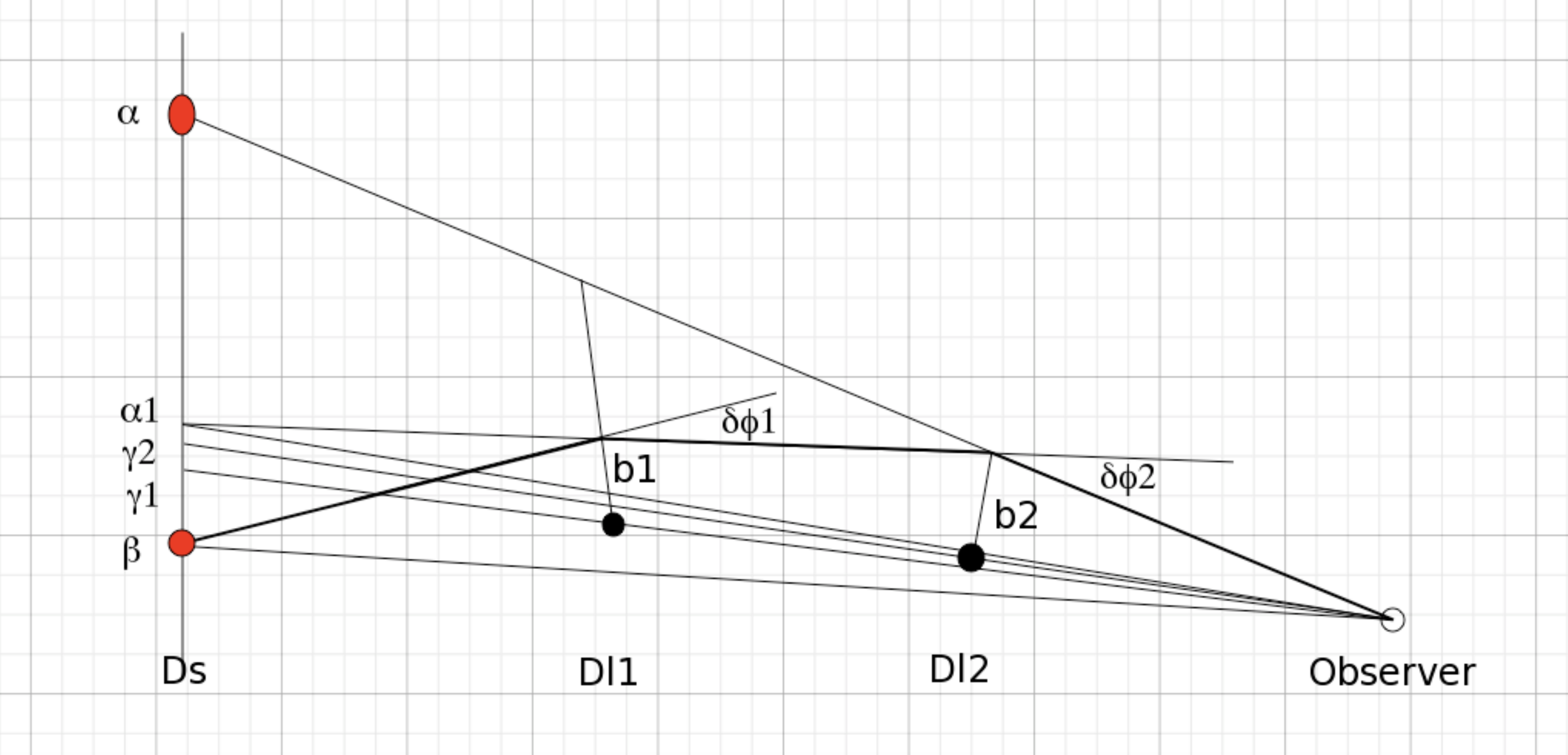} 
\caption{
A diagram of a photon path scattered by lenses 1 and 2 at distances
$D_{l1}$ and $D_{l2}$ in sequence. 
The scattering planes are noncoplanar in general, and the angles
should be considered as two-vectors in three space. 
} 
\label{fig_double_ray} 
\end{figure}

The double scattering two point lens equation can be obtained from
a diagram shown in figure \ref{fig_double_ray} where the linear gravity 
and small angle approximation are assumed. 
Since the true photon path is three dimensional
because of the rotation of the scattering planes with respect to each 
other, a three-dimensional diagram is needed. But it has been 
shown in RB10 that in the linear approximation in small angles,
the radial component (in the direction of the line of sight) of the
impact vector that is generated due to the relative rotation of the 
scattering planes can be ignored because it is of the second order.   
It is sufficient to express the triangular relations of the angles
in vectors to account for the relative rotaion between the two scattering 
planes. 

From figure \ref{fig_double_ray} two sets of relations are obtained.
\opeqn
  \vec b_1 = D_{l1} (\vec\alpha - \vec\gamma_1)
     + (D_{l1} - D_{l2}) \vec\delta\varphi_2 \ ; \qquad
  D_s (\vec\alpha_1 - \vec\beta) = - \vec\delta\varphi_1 (D_s - D_{l1})
\label{eqFirstScatter}
\cleqn
\opeqn
  \vec b_2 = D_{l2} (\vec\alpha - \vec\gamma_2) \ ; \qquad
  D_s (\vec\alpha - \vec\alpha_1) = - \vec\delta\varphi_2 (D_s - D_{l2})
\label{eqSecondScatter}
\cleqn
where
the bending (scattering) angles are given by the point mass bending angles.  
\opeqn 
 \vec\delta\varphi_1 = \frac{4GM_1 (-\hat b_1)}{b_1} \ ; \qquad
\vec\delta\varphi_2 = \frac{4GM_2 (-\hat b_2)}{b_2} \ 
\label{eqBending}
\cleqn
$M_1$ and $M_2$ are the masses of the first and second point mass scatterers
at the distances $D_{l1}$ and $D_{l2}$, and $D_s$ is the distance to the 
source; $\vec b_1$ and $\vec b_2$ are the
impact vectors, and $b_1 \equiv |\vec b_1|$ and $b_2 \equiv |\vec b_2|$. 
The lens equation is obtained from the second equations 
of eqs. (\ref{eqFirstScatter}) and (\ref{eqSecondScatter}),
\opeqn
 D_s (\vec\alpha - \vec\beta) = - \vec\delta\varphi_1 (D_s - D_{l1})
    - \vec\delta\varphi_2 (D_s - D_{l2})\,, 
\label{eqPreLensEq}
\cleqn
which is completed by using eq.(\ref{eqBending}) and the first equations of
eqs. (\ref{eqFirstScatter}) and (\ref{eqSecondScatter}).

It is convenient (or our custom) to define a lens plane and use 
linear variables instead of the angular variables. 
Note that the intermediate image position angle
$\vec\alpha_1$ was defined by projecting the intermediate 
photon ray back to the sky at the distance of the source.
So define the lens plane, where the lens equation variables are defined,
as the plane at the distance of the source and normal to a chosen radial 
direction. The lens equation is independent of the choice of the radial 
direction because of the linear approximation in small angle.
Since the lens plane is placed at the distance of the source, 
the linear variables are $D_s$ times the angular variables.

Now employ the complex coordinates as usual and let $\omega$, $z$, 
and $x_j: j=1,2$ denote the (2-dimensional) positions (on the lens
plane at the distance of the source) of a source, an image, and
lenses $1$ and $2$. Then the lens equation in eq.(\ref{eqPreLensEq})  
can be written in terms of the linear variables.
\opeqn
 \omega = z - \frac{r_{E1}^2}{\bar z_1 -{r_{E21}^2}z_2^{-1}} 
            - \frac{r_{E2}^2}{\bar z_2} \,, 
\label{eqUnnormLeq}
\cleqn
where $ z_j \equiv z - x_j: \ j=1,2$. 
Let $r_{Ekj}$ be the (single lens)  Einstein ring radius of 
the lensing by object $k$ of object $j$, and let object $0$ 
refer to the source. 
$r_{E1}\equiv r_{E10}$, $r_{E2}\equiv r_{E20}$, and $r_{E21}$ are
as follows.
\opeqn
  r_{E1}^2 = 4GM_1D_1 \frac{D_s^2}{D_{l1}^2} = R_{E1}^2\frac{D_s^2}{D_{l1}^2} 
   ; \qquad  D_1 \equiv \frac{D_{l1}(D_s - D_{l1})}{D_s}
\cleqn
\opeqn
  r_{E2}^2 = 4GM_2D_2 \frac{D_s^2}{D_{l2}^2} = R_{E2}^2\frac{D_s^2}{D_{l2}^2} 
   ; \qquad  D_2 \equiv \frac{D_{l2}(D_s - D_{l2})}{D_s}
\cleqn
\opeqn
  r_{E21}^2 = 4GM_1D_3 \frac{D_s^2}{D_{l2}^2} = R_{E21}^2\frac{D_s^2}{D_{l2}^2}
   ; \qquad  D_3 \equiv \frac{D_{l2}(D_{l1} - D_{l2})}{D_{l1}}
\cleqn
$D_j: j =1, 2, 3$ are the reduced distances, and $R_{Ekj}$ is the 
``intrinsic" Einstein ring radius of the lensing of object $j$ by 
object $k$. The reason why we refer to $R_{Ekj}$ as the intrinsic Einstein ring
radius is that the photon rays of the Einstein ring image of the lensing
(of object $j$ by object $k$) actually pass through the ring 
around the lens ($k$) of radius $R_{Ekj}$ (accurate within the
small angle approximation).

Redefine distances $D_{j1} \equiv D_{lj}$ and $D_{j2} \equiv D_s - D_{lj}$
and define effective masses 
\opeqn
effM_j \equiv M_j \frac{D_{j2}}{D_{j1}} \ : \qquad j=1,2 \ . 
\cleqn 
Define the Einstein ring radius $r_E$ of the total effecitve mass,
\opeqn
 r_E^2 \equiv r_{E1}^2 + r_{E2}^2 = 4 G D_s (effM_1 + effM_2) \,, 
\cleqn
and the lens equation can be normalized so that the unit distance is $r_E$. 
By substituing $\omega$, $z$ and $x_j$ in eq.(\ref{eqUnnormLeq}) by 
$r_E\omega$, $r_E z$ and $r_E x_j$ respectively, the normalized lens
equation is obtained. 
\opeqn
\omega = z - \dfrac{\epsilon_1}{\bar z_1 - \dfrac{a}{z_2}}
            - \frac{\epsilon_2}{\bar z_2} \,,
\label{eqLeq}
\cleqn
where the effective fractional masses are 
\begin{eqnarray}
 \epsilon_1 &\equiv \dfrac{r_{E1}^2}{r_E^2} 
             = \dfrac{effM_1}{effM_1 + effM_2}
             = \dfrac{M_1}{M_1 + M_2/d}  
             = \dfrac{\epsilon}{1 + \epsilon}    \\ 
 \epsilon_2 &\equiv \dfrac{r_{E2}^2}{r_E^2}
             = \dfrac{effM_2}{effM_1 + effM_2}
             = \dfrac{M_2}{M_1 d + M_2}   
             = \dfrac{1}{1 + \epsilon}
\end{eqnarray}
and the double scattering parameter is 
\opeqn
  a \equiv \dfrac{r_{E21}^2}{r_E^2}
    = \dfrac{M_2(1-d)}{M_1 d + M_2}
    = \frac{1-d}{1+\epsilon}
\cleqn
The distance parameter $d$ is
\opeqn
 d \equiv \frac{D_{12}D_{21}}{D_{11}D_{22}}\ \  \leq \ 1  
\cleqn
where the equality in the second relation holds when the two
lenses are at the distance.
The effective mass ratio $\epsilon$ is
\opeqn
 \epsilon \equiv \frac{\epsilon_1}{\epsilon_2}
          = \frac{r_{E1}^2}{r_{E2}^2}
          = \frac{effM_1}{effM_2}
          = d \frac{M_1}{M_2} 
\cleqn
The effective mass ratio is smaller than the mass ratio.
The weight is shifted to the last scatterer in double scattering lensing.
Note that we have chosen the last scatterer for the reference mass. 

It should be worth pointing out that the double scattering parameter $a$  
is essentially the (square of the) Einstein radius that is easily measurable   
in an (almost) axisymmetric system as in SDSSJ0946+1006 
\citep{double-ring-first}.  In an axisymmetric DSTP lens
three ring images are formed, even though the innermost ring is ``unstable"
to break into a half-circle, and $\sqrt{a}$ measures the middle ring radius
in units of the Einstein ring radius of the total effective mass $r_E$. 
The DSTP lens system can be considered to have two characteristic parameters 
$r_E$ and $\sqrt{a}$.

Here the focus is in the main lens and
the interest is on what happens to the Einsteing ring of the main lens 
under the perturbation of a perturbing mass. So it is useful to 
renormalize the lens equation by the Einstein ring radius of the main lens.
There are two cases: 1) object 1 is the perturbing mass; 
2) object 2 is the perturbing mass. 

Case 1): 
Renormalize the lens equation (\ref{eqLeq}) so that the unit distance
is $r_{E2}$.
\opeqn
 \omega = z - \frac{\epsilon}{\bar z_1 - \tilde a_2 z_2^{-1}}
            - \frac{1}{\bar z_2}  
\label{eqLeqOne}
\cleqn
where
\opeqn
 \tilde a_2 \equiv \frac{r_{E21}^2}{r_{E2}^2} = 1-d 
\cleqn

Case 2): 
Renormalize the lens equation so that the unit length is $r_{E1}$. 
\opeqn
  \omega = z - \frac{1}{\bar z_1 - \tilde a_1 z_2^{-1}} 
        - \frac{\epsilon^{-1}}{\bar z_2}
\label{eqLeqTwo}
\cleqn
where
\opeqn
 \tilde a_1 \equiv \frac{r_{E21}^2}{r_{E1}^2} = \frac{1-d}{\epsilon}
\cleqn

\section{Large Separation DSTP Lenses: $\ell >> 1$}

\subsection{When the Perturber is the First Scatterer}

\subsubsection{The Lens Equation}

Let the separation be denoted by $\ell \equiv |x_1 - x_2|$.  
The coordinate system can be chosen such that lens 2 
is at the origin, $x_2 =0$, and lens 1 is on the positive side 
of the real axis, $x_1 = \ell$. 
Since it is assumed that $\ell >> 1$, the lens equation (\ref{eqLeqOne})
can be expanded in power series in $\ell^{-1}$, assuming that $\epsilon$ is 
not bigger than ${\cal O}(1)$, to obtain the following.
\opeqn
 \omega -\frac{\epsilon}{\ell} 
     = z - \frac{1}{\bar z}
         + \frac{\epsilon}{\ell^2}\bar z 
         - \frac{\epsilon }{\ell^2}\frac{\tilde a_2}{z}
\label{eqApproxLeqOne}
\cleqn
The lens is made of a point mass ($\propto 1/\bar z$), a constant shear 
($\propto \bar z$), and a mass-antimass distribution($\propto 1/z$); 
the source is shifted by $\epsilon/\ell$ as is the case with the wide
separation binary lens. (See appendix.) 
Consider the RHS minus the LHS 
as a vector field. It is a vector field with zeros and poles on 
the two sphere, and the index of the vector field 
at $z \sim\infty$ results in $n_+ - n_- = 0$ where $n_+$ and $n_-$ are the 
number of positive and negative images respectively. 
(See RB10.) Thus, the number
of images is even and the number of negative images is the same as the
number of positive images. There are two images for $\omega = \infty$, namely
$z = 0$ and $\infty$, hence there are two or four images where the latter
occurs inside the finite size caustic. The finite size caustic occurs
because the degeneracy of the point caustic of the single lens is 
broken by the perturbation of the mass $M_1$. The size of the caustic
curve is calculated below using second order approximation in $\ell^{-1}$.

\subsubsection{The Critical Curve and Caustic Curve}

The Jacobian of the lens equation is given as 
\opeqn
  Jacobian = \pmatrix{f \  g \cr
                      \bar g \ \bar f} \ :
\cleqn
\opeqn
  f \equiv \partial \omega 
    = 1 + \frac{\epsilon }{\ell^2} \frac{\tilde a_2}{z^2} 
    ; \qquad  
  g \equiv \bar \partial \omega 
    = \frac{\epsilon }{\ell^2} + \frac{1}{\bar z^2} .
\cleqn
where $\partial \equiv \partial/\partial z$ 
and $\bar\partial \equiv \partial/\bar\partial z$.
The Jacobian determinant is 
\opeqn
  J = |f|^2 - |g|^2
\cleqn
and the eigenvalues of the Jacobian are 

\opeqn
  \lambda_{\pm} = f_R \pm \left(|g|^2-f_I^2 \right)^{1/2}
\cleqn 
where $f_R$ and $f_I$ are the real and imaginary parts of $f$.
On the critical curve, where $J =0$, one or both of the eigenvalues 
are zero because $J$ is the product of the eigenvalues.
\opeqn
  \lambda_\pm = f_R \pm f_R  
\cleqn
Thus $\lambda_-$ vanishes on the critical curve, and $\lambda_+$
also vanishes if $f_R = 0$. Note that $f = f_R = 1$ in the case of 
the binary lens, and $\lambda_+$ never vanishes. Here $f_R > 0$ 
because $\ell >> 1$.

The lens system is simple enough so that the 
critical curve can be explicitly written out as a simple function.
If we set $z = r e^{i\theta}$, the critical condition is given 
by the following in the linear approximation in $\epsilon/\ell^2$.
\opeqn
  r = 1 + \frac{\epsilon d}{2\ell^2}\cos 2\theta
\label{eqCriticalCurveOne}
\cleqn
Compared to the circular critical curve $r=1$ of the main (single)
lens, the critical curve is slightly squeezed in a quadrupolar
manner. Note that every point of the entire ring $r=1$ of the 
single lens is a precusp ({\it i.e.,} mapped to a cusp). 
The curve in eq.(\ref{eqCriticalCurveOne}) is circular 
($dr/d\theta = 0$) at four points: $0$, $\pi/2$, $\pi$, and $3\pi/2$,
and they are expected to be the precusps. It is the indeed the case
as will be shown shortly. The size of the caustic can be estimated by 
calculating the cusp positions using the lens equation.   
The precusps along the real axis, $\theta = 0$ and $\pi$, 
are mapped to cusp points on the 
real axis, and the length of the caustic along the real axis is 
obtained as the difference between the cusp positions.
\opeqn
 \Delta \omega_{real} =  \omega(x) - \omega(-x) 
  = \frac{4\epsilon d}{\ell^2} 
; \qquad
    x = 1 + \frac{\epsilon d}{2\ell^2} 
\label{eqDeltaOmegaX}
\cleqn
The length of the caustic in the direction parallel to the imaginary 
axis is given as the absolute value of the following.
\opeqn
 \Delta \omega_{imag} =  \omega(iy) - \omega(-iy) 
  = -i \frac{4\epsilon d}{\ell^2} 
; \qquad
    y = 1 - \frac{\epsilon d}{2\ell^2}
\label{DeltaOmegaY}
\cleqn 
\opeqn
 |\Delta \omega_{real}| = |\Delta \omega_{imag}| 
   = \frac{4}{\ell^2}\frac{M_1}{M_2} d^2
\cleqn
Thus the quadroid caustic is equilateral 
and the orientation of the caustic is opposite to the critical curve.
In comparison to the wide separation binary lens, for which $d=1$,
the size of the caustic is smaller by factor $d^2$. See the Appendix
for the wide separation binary. If the lens 
elements are evenly distributed in distance between the observer and
the source, then $d = 1/4$, and the caustic shrinks by $1/16$. 
It is substantial, and it demonstrates that perturbation of microlensing 
events by a ``rogue" mass in an angular proximity 
should be estimated by using a proper double scattering lens 
equation. It has been the practice that all possible perturbers are  
universally thrown into constant shear corrections, or constant shear
and constant convergence.

\subsubsection{Cusps}

On the critical curve, $f_R > 0$, hence $\lambda_- = 0$ is
responsible for $J =0$. Thus, if $e_+$ and $e_-$ are the eigenvectors
coresponding to $\lambda_+$ and $\lambda_-$ respectivley, then 
$e_-$ is the critical direction. If we consider drawing the caustic curve
by mapping the critical curve by the lens equation, 
only the non-critical ($e_+$) component of 
the tangent vector of the critical curve is mapped to the tangent of 
the caustic curve because of the criticality condition. Thus the caustic
curve is tangent to the eigendirection of $\lambda_+$ \citep{rhie99,rhie01}.
If the tangent to the critical curve is parallel to the critical direction,
the tangent mapped to the caustic curve is zero and the progression of the   
caustic curve stops and forms a cusp. 
In the next moment, the non-critical component is 
picked up and the caustic curve turns around changing the direction by 
$\pi$.  
If $p$ is the parameter of the critical curve, the tangent to the curve
is determined by
\opeqn
  0 = \frac{dJ}{dp} = \frac{dz_+}{dp}\partial_+J 
                      + \frac{dz_-}{dp} \partial_-J 
\cleqn
where $dz_\pm$ are the increments in the $\pm$ eigendirections.
The cusp forms when $dz_+/dp = 0$, hence $0 = \partial_-J$. 
The (unnormalized) eigenvectors are  
\opeqn
  e_+ \propto \pmatrix{ u_+ \cr
                  v_+} ; \qquad
  e_- \propto \pmatrix{ u_- \cr
                  v_-},
\cleqn
where we can choose $u_\pm$ and $v_\pm$ as 
\begin{eqnarray}
  & u_+ = g \ ; \qquad 
  v_+ = - i f_I + \left(|g|^2-f_I^2 \right)^{1/2}  \cr
  & u_- = g \ ; \qquad 
  v_- = - i f_I - \left(|g|^2-f_I^2 \right)^{1/2}
\label{eqUV}
\end{eqnarray}
The Jacobian matrix can be diagnolized using $\Lambda$ constructed
from the eigenvectors components,
\opeqn
  \Lambda = \pmatrix{u_+ \ u_- \cr
                     v_+ \ v_-} : \qquad
  \Lambda^{-1} (Jacobian) \Lambda 
  = \pmatrix{\lambda_+ \ 0 \cr
             0  \ \lambda_- }, 
\cleqn
and the eigendirection differentials can be written as 
\opeqn
  \pmatrix{dz_+ \cr
           dz_-} 
    = constant\, \Lambda^{-1} \pmatrix{dz \cr d \bar z}
\cleqn 
where $constant$ is a real constant.
Thence $\partial_- z = constant\ u_-$ and and 
$\partial_-\bar z = constant\ v_-$,
and the cusps are found from the cusp condition.
\opeqn
 0 = \partial_-J = \partial_- z \partial J + \partial_- \bar z \bar \partial J
  \qquad \Rightarrow \qquad
   0 = u_-  \partial J + v_- \bar \partial J
\cleqn
Straightforward calculations show that, in the second order 
in $\ell^{-1}$, $\partial_-J=0$ for $\theta = 0$, $\pi/2$, $\pi$, and $3\pi/2$ 
of the critical curve in eq.(\ref{eqCriticalCurveOne}).
Therefore they are precusps in the second order approximation.

\subsubsection{Images}

\begin{figure}
\plotone{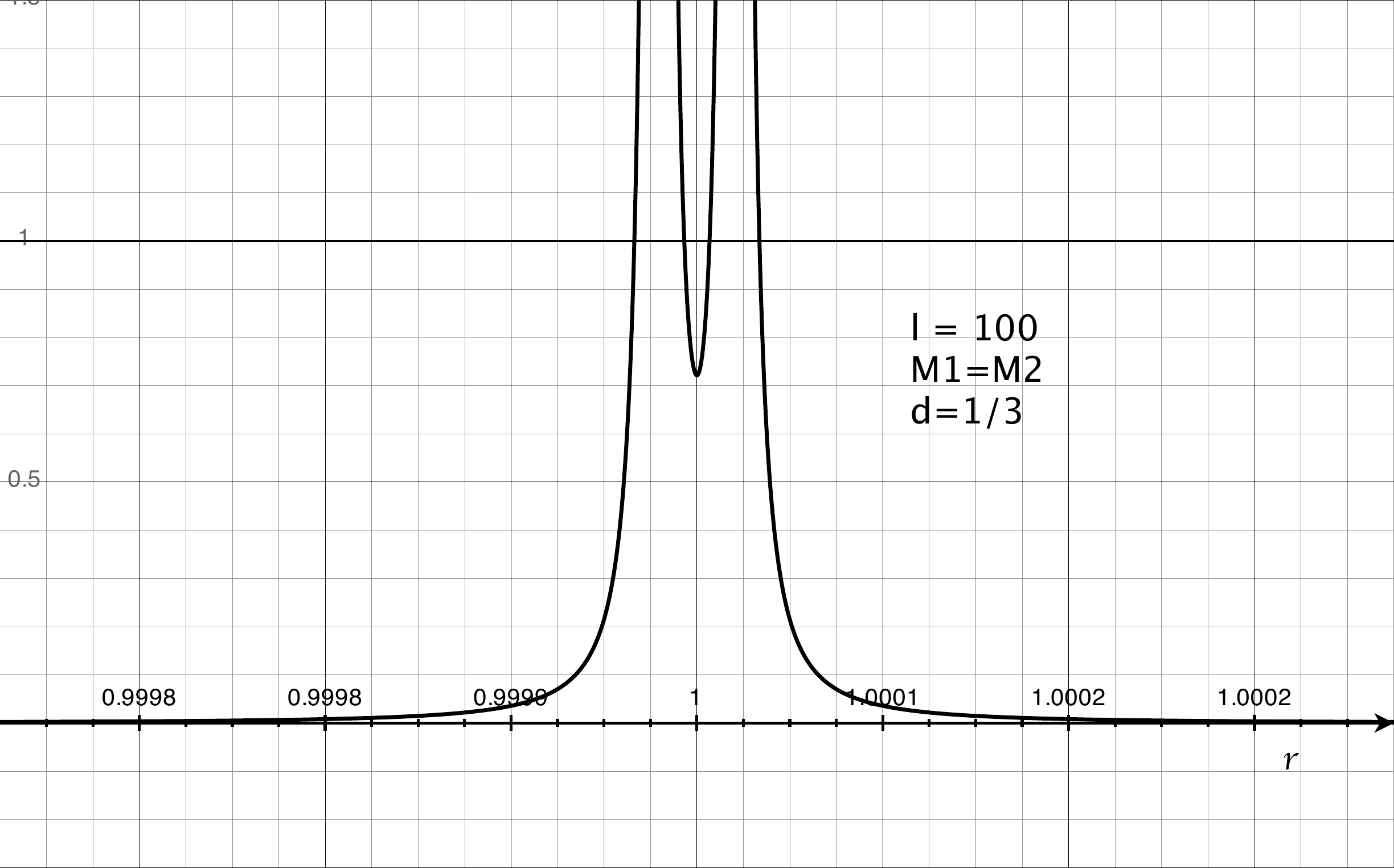}
\caption{The radii of the four images of a source 
at $\zeta_1=\zeta_2=2\times 10^{-5}$. They are all near 1 -- 
the Einstein ring radius of the main lens $M_2$.
The lens parameters are $\ell =100$, $d=1/3$, and $M_1 = M_2$. 
}
\label{fig_leqone}
\end{figure}

Set $z = r e^{i\theta}$ and $\zeta =\omega - \epsilon/\ell$. 
The lens equation for the shifted source position is given by
\opeqn
  \zeta = \xi e^{i\theta} + \eta e^{-i\theta}  
\cleqn
where $\xi$ and $\eta$ are functions of $r$. 
\opeqn
  \xi = r - \frac{1}{r} ; \qquad
  \eta = \frac{\epsilon}{\ell^2}\left(r - \frac{\tilde a_1}{r}\right)
\cleqn
If we let $\zeta = \zeta_1 + i \zeta_2$,
\opeqn
  \zeta_1 = (\xi + \eta)  \cos\theta ; \qquad
  \zeta_2 = (\xi - \eta)  \sin\theta ,
\label{eqImageAngle}
\cleqn
and an equation for  $r$ is obained.
\opeqn
   \left(\frac{\zeta_1}{\xi+\eta}\right)^2 
     + \left(\frac{\zeta_2}{\xi-\eta}\right)^2 = 1
\label{eqRadialOne}
\cleqn
There are two or four solutions to the equation (\ref{eqRadialOne}),
which indicates that there are two solutions outside the quadroid
caustic and four solutions inside the caustic. When the source is 
inside the caustic, the images are all at $r \approx 1$. They are
the four bright images that form around the critical curve $r\approx 1$,
two outside the critical curve in the area of the ``squeezed" and   
two inisde the critical curve in the area of the ``bulged".
For example, $\zeta =0$ is inside the caustic, and the four images
are on the real axis and the imaginary axis. 
\opeqn
  r_{1,2}^2 = 1 +\frac{\epsilon}{\ell^2}(1 - \tilde a_2) 
          \ ; \qquad 
   \theta_1 = \frac{\pi}{2}, \quad \theta_2 = \frac{3\pi}{2}
\cleqn
\opeqn
  r_{3,4}^2 = 1 -\frac{\epsilon}{\ell^2}(1 - \tilde a_2) 
     \ ; \qquad 
   \theta_3 = 0, \quad \theta_4 = \pi 
\cleqn
The radius $r_{1,2}$ is bigger than $r_{3,4}$ because $\tilde a_2 < 1$
for a double scattering lens with $d < 1$.
Generally, the radii of the images are different.
Figure \ref{fig_leqone} shows a case: $\ell=100$, $d=1/3$, and $M_1=M_2$;
$\zeta_1 = \zeta_2 = 2.\times 10^{-5}$. 
The angle $\theta$ for the radius $r$ of each image is determined
from eq.(\ref{eqImageAngle}).  
The four images of a finite size source filling the caustic form 
more or less a circular ring with finite thickness threaded by $r=1$.

\subsection{When the Perturber is the Last Scatterer}

\subsubsection{The Lens Equation}

Power-expand eq.(\ref{eqLeqTwo}) in $\ell^{-1}$ assuming
$|z| \approx 1$ because we are interested in the region around
the critical curve. Keep the terms up to the second order in $\ell^{-1}$ 
because the size of the caustic is of the second order. It will be shown
that the linear order perturbation shifts the position of the caustic,
by $d(\epsilon \ell)^{-1}$, 
but does not break the degeneracy of the point caustic.
\opeqn
 \omega -\frac{1}{\epsilon\ell}
    = z - \frac{1}{\bar z} + \frac{\tilde a_1}{\ell \bar z^2}
        + \frac{\tilde a_1 z}{\ell^2 \bar z^2}
        - \frac{\tilde a_1^2}{\ell^2 \bar z^3}
        + \frac{\bar z}{\epsilon\ell^2}
\label{eqApproxLeqTwo}
\cleqn
The lens is made of a point mass, a constant shear, and a whole
variety of multipoles. From the index of the vector field with zeros
and poles, it is obtained that $n_+-n_-=-2$ where $n_\pm$ is the 
number of positive/negative images. Thus the number of images is even. 
For $\omega = \infty$, there are four images, 
one at $z=\infty$ and three degenerate images at $z=0$. As $\omega$
moves toward the lenses, the three degenerate images individualizes.
It is expected that the number of images is four outside the
caustic and six inside so that $n_+-n_-=-2$.  It is known that the
original lens equation (eq.(\ref{eqLeqOne}) or (\ref{eqLeqTwo}))
without approximations produces four or six images 
\citep{ES93, petters}, and RB10 succeeded in deriving the sixth order
analytic polynomial equation from the lens equaiton. However, the 
approximate lens equation (\ref{eqApproxLeqTwo})
has a third order pole, and we will see that
two images are dark images remaining  near the pole. We refer
to them as ignorable images.

\subsubsection{The Critical Curve and Caustic}

Jacobian matrix components are
\opeqn
 f = \partial \omega = 1 + \frac{\tilde a_1}{\ell^2 \bar z^2} ; \qquad 
 g = \bar\partial\omega 
   = \frac{1}{\bar z^2} - \frac{2\tilde a_1}{\ell \bar z^3}
        - \frac{2\tilde a_1 z}{\ell^2 \bar z^3}
        + \frac{3\tilde a_1^2}{\ell^2 \bar z^4}
        + \frac{1}{\epsilon\ell^2} .
\cleqn
Set $z = r e^{i\theta}$ and the Jacobian determinant 
$J = |f|^2 -|g|^2$ can be computed up to the second order. 
\opeqn
 J = 1 - \frac{1}{r^4} 
      + \frac{4 \tilde a_1 \cos\theta}{\ell r^5}
     +
      \frac{2 \tilde a_1 \cos\theta}{\ell^2 } 
        \left(\frac{1}{r^2} - \frac{2}{r^4} \right)
       - \frac{2\cos\theta}{\epsilon\ell^2 r^2}
       - \frac{4 \tilde a_1^2}{\ell^2 r^6}
       - \frac{6 \tilde a_1^2 \cos 2\theta}{\ell^2 r^6}
\label{eqJTwo}
\cleqn
In the linear order,   
\opeqn
 J_{linear}
 = 1 - \frac{1}{r^4}\left(1 - \frac{4\tilde a_1 \cos\theta}{\ell r} \right),
\cleqn
and the critical curve, $J_{linear}=0$, is given by 
\opeqn
 \frac{4\tilde a_1 \cos\theta}{\ell} = r - r^5 ; \qquad r \neq 0 .
\cleqn
Using graph of the RHS and the fact that $|\cos\theta| \leq 1$,
it is found that the solution space is near $r=0$ and $r=1$.
We are interested in the critical curve with $r\approx 1$. Let 
$r = 1 - \delta$ and  obtain the critical curve in the linear
order in $\delta$. 
\opeqn
  r = 1 - \frac{\tilde a_1 \cos\theta}{\ell}. 
\cleqn 
It is a cardioid even though it is hard to distinguish from a 
circle because of the small coefficient of the $\cos\theta$ term. 
The whole critical curve is mapped by the approximate lens equation 
(\ref{eqApproxLeqTwo}) to a point 
$\omega = d(\epsilon\ell)^{-1}$, hence the caustic is a point caustic
in the linear order. It is shifted from that of the single lens.
The position is different by factor $d$ from the center of the caustic 
of the large separation binary lens which is $(\epsilon\ell)^{-1}$.

We need the second order, and the second order depends on $\cos2\theta$.
The combination of $\cos\theta$ terms and $\cos 2\theta$ terms
produces an ``egg-shape" curve that resembles the familiar
quaroid critical curve of the binary lens.  
Set $r = 1 - \delta$ and
\opeqn
 \delta = \frac{A}{\ell} + \frac{B}{\ell^2} 
\cleqn 
and find $A$ and $B$ from the full Jacobian determinant in 
eq.(\ref{eqJTwo}). 
\opeqn
A = \tilde a_1 \cos\theta \ ; \qquad
B = \frac{3\tilde a_1}{2}\cos 2\theta 
     - \frac{1}{2\epsilon}\cos 2\theta
     + \frac{\tilde a_1^2}{4}(1 - \cos 2\theta)
\cleqn 
The critical curve $r(\theta)$ is a linear function of $\cos\theta$
and $\cos 2\theta$, hence it has the shape of an asymmetric peanut 
(squeezed in at $\theta=0$ and $\pi$) or a pear depending on whether
the coefficient of $\cos 2\theta$ is negative or positive. (The 
coefficient of $\cos\theta$ is negative.)  

For $\theta = 0$ and $\pi$, $dr/d\theta = 0$, and they are suspected
to be precusps. They are, as can be shown by calculating $\partial_- J$ 
as was done for case 1). The other two precusps can be seen numerically
to occur not exactly but practically at $\theta = \pi/2$ and $3\pi/2$.   
The cusps on the real axis will be
used to estimate the size of the caustic. 
\opeqn
  \Delta \omega_{real} = \omega_0 - \omega_{\pi} 
   = \frac{4d}{\epsilon\ell^2} 
   = \frac{4}{\ell^2}\frac{M_2}{M_1}
\cleqn 
It is of order $1/\ell^2$ and is the same size as the large separation
binary lens. But the shape of the critical curve is different from that
of a binary lens as was mentioned above. Such a nice result should have
a physical interpretation which escapes our mind currenlty. It should
be worth pondering in an idle time.

The majority of the
microlensing toward the Galactic bulge is bulge-bulge lensing and 
the microlensing event can be perturbed by a foregroud star. 
The highest magnification microlensing
event observed to date is of the total magnification 2400, which 
means that the impact distance is $4.17\times 10^{-4}$ Einstein ring
radius. If $M_1=M_2$ and $\ell=100$, then the ``radius" of the caustic
is $2\times 10^{-4}$. If the main lens has solar mass and is at 
$D_{l1}= 3 D_s/4$, the Einstein ring radius is $r_{E1} = 2310 sec$. 
If the source star is a sun-like star with the solar radius ($2.32$ sec), 
it is $10^{-3}$ in units of the Einstein ring radius.
The caustic is completely inside the source star radius. 
Thus perturbers with $\ell \lsim 1/100$ are expected to contribute to 
measurable effects. Of course, a more massive perturber (large $M_2$)
perturbs more strongly, and when it is much larger than the main lens,
the calculations done here are not proper because we would need higher
order terms.

\subsubsection{Cusps}

In order to see that the critical points given by eq.(\ref{eqJTwo})
are precusps where $\theta = 0$ and $\pi$, calculate 
$\partial_-J$ for $\theta = $ and $\pi$. 
$\partial_- J = u_- \partial J + v_- \bar\partial J$ where 
$u_-$ and $v_-$ are given in eq.(\ref{eqUV}).
It is easy to see that for $\theta = 0$ and $\pi$, 
$v_- = -|g| = -g = u_-$, 
and $\partial J = \bar\partial J$ in the second order in $1/\ell$. 
Therefore, $\partial_- J = 0$, and the points with $\theta = $ and $\pi$
are precusps.

\subsubsection{Images and Ignorables}

In the linear order, the lens equation reads as follows.
\opeqn
 \omega -\frac{1}{\epsilon\ell}
   = z - \frac{1}{\bar z} + \frac{\tilde a_1}{\ell \bar z^2}
\cleqn
It produces odd number of images
with one more negative image than the positive image 
as one can see from the pole at $z=\infty$ and a double pole at $z=0$:
$n_+ -n_-=-1$. For $\omega=\infty$, there are three images, 
one at $z=\infty$ and two degenerate images at $z=0$. 
Let's look at the images at $\omega = 1/(\epsilon\ell)$ which 
can be solved easily.  As was discussed before, the point caustic is at 
$\omega = d(\epsilon\ell)^{-1}$, and $\omega = 1/(\epsilon\ell)$ is
outside the point caustic and generates three images.
\opeqn
  z = \frac{\tilde a_1}{\ell} \ , \qquad
      1-\frac{\tilde a_1}{2\ell} \ , \qquad
      -1 - \frac{\tilde a_1}{2\ell} \qquad
\cleqn 
The second image is the positive image located outside the critical 
curve where it is squeezed in. The third image is a negative image
located inside the critical curve where it is bulged out. The first
image is the dim image located near the lens position $z=0$. 
The first and third images become degenerate when $\omega\rightarrow \infty$.
One can see that the third image moves fast and the first image slowly.
Because the caustic is a point caustic, all the three image trajectories 
are continuous. Physically, the first image is ignorable, and practically
there are two images as is the case in the single lens.   

The full lens equation (in the second order) in eq.(\ref{eqApproxLeqTwo})
has a triple pole at the main lens position $z=0$ which are image 
positions of $\omega = \infty$. Two of them are 
ignorable because they are confined to the very close proximity to 
the lens position and have negligible fluxes for finite $\omega$. 
In order to find the 
approximate positions of the ignorable images for small $\omega$ 
(in the near proximity of the caustic), set $z = A/\ell$ in the lens
equation (\ref{eqApproxLeqTwo})
and find $A$ for which the large terms (depending on the 
positive power of $\ell$) add to zero. There are two solutions.
\opeqn
  z = \frac{\tilde a_1}{2\ell}(1\pm i\sqrt{3})
\cleqn    
The position of the corresponding source is ${\cal O}(1/\ell)$.  
Therefore there are two or four images in practice.

The lens equation (\ref{eqApproxLeqTwo}) can be converted into an 
analytic equation and it is a fifth order equation when truncated 
in the second order in $1/\ell$. One of the solutions is an ignorable
image. The caustic is centered at $\omega \sim d(\epsilon\ell)^{-1}$.

\section{The Double Scattering Two Distributed Mass Lens and
            Large Separation Approximation}

A galaxy lens (of finite mass and finite extension) can be expressed 
in terms of its projected mass density $M\sigma$ where $M$ is the
total mass and $\sigma$ is the normalized projected mass density.
\opeqn
 \int \sigma = 1
\cleqn
In the case of a point mass, $\sigma$ is the (Dirac) delta function.
If $M_1\sigma_1$ and $M_2\sigma_2$ are the mass densities of the 
galaxy lenses 1 and 2, the double scattering two distributed mass 
(DSTD) lens equation is written as follows where the unit distance
is given by $r_E$ the Einstein ring radius of the 
total (effective) mass.     
\opeqn
\omega = z - \epsilon_1 \dint 
              \dfrac{d^2 x^\prime \sigma_1(x^\prime)}
               {\bar z_1 - \bar x^\prime 
               - a \dint  
               \dfrac {d^2 x^{\prime\prime} \sigma_2(x^{\prime\prime})}
                      {z_2 - x^{\prime\prime}}}
           - \epsilon_2 \int
              \dfrac{d^2 x^\prime \sigma_2(x^\prime)}
               {\bar z_2 - \bar x^\prime}
\label{eqDSTD}
\cleqn
where $z_j \equiv z - x_j: j = 1, 2$ as before and 
$x_j$ is the center of mass position of the $j$-th galaxy lens. 
The density functions $\sigma_j$ are real valued functions and
$d^2 x^\prime$ is the real 2-d volume element: 
$d^2 x^\prime = dx^\prime\wedge d \bar x^\prime /(-2i)$.
 
If galaxy 1 is the perturber, we can set $x_1 = \ell$ and $x_2 =0$,
and assume that $\ell >> 1$.
Again here we are interested in the neighborhood of the critical curve
where the bright (or detectable) images of a quasar or a galaxy of a 
high redshift are found. Since the perturbed critical curve would be 
a small modification of the critical curve of the unperturbed lens, 
we renormalize the lens equation so that the critical curve would be
given by $|z| \approx 1$.  
Let's for conenience denote the second deflection angle 
(multiplied by a distance factor and normalized) by $\Pi_2$. 
\opeqn
 \Pi_2 (z;x_2) \equiv \int
              \dfrac{d^2 x^\prime \sigma_2(x^\prime)}
               {\bar z -\bar x_2 - \bar x^\prime}
\cleqn
The equation (\ref{eqDSTD}) can be rewritten as follows where
the unit distance is given by $r_{E2}$.
\opeqn
 \omega = z - \epsilon\int\dfrac{d^2 x^\prime \sigma_1(x^\prime)}
              {\bar z - \ell - \bar x^\prime - \tilde a_2 \overline{\Pi_2(z;0)}} 
            - \Pi_2(z;0)
\label{eqDSTDApproxOne}
\cleqn
where $x_2 =0$ and 
the double scattering parameter $\tilde a_2$ is defined exactly the same way
as in the case of the point mass lenses but with the distances of the 
center of the masses.
Now the critical curve is in the neighborhood of $|z|=1 $,   
and $\Pi_2$ can be assumed to be ${\cal O}(1)$. 

The equation (\ref{eqDSTDApproxOne}) can be power-expanded in $1/\ell$
to obtain 
\opeqn
 \omega - \frac{\epsilon}{\ell} 
      = z - \Pi_2(z;0) 
           + \frac{\epsilon}{\ell^2}\bar z
           - \frac{\epsilon \tilde a_2}{\ell^2}\overline{\Pi_2(z;0)}
\label{eqGalaxyOne}
\cleqn
Since the dipole moment of the mass distribution $\sigma_1$ is zero,
the perturbing galaxy 1 behaves as a point mass lens.
The last term of eq.(\ref{eqGalaxyOne})
depends on the double scattering parameter $\tilde a_2$
and vanishes for a perturbing galaxy at the same distance as the main
lens leaving only the constant shear term. 
It would be  an error to ignore the last term if the (first scattering) 
perturbing galaxy is at a different distance.

If the perturbing galaxy is lens 2, we can set $x_1 =0$ and $x_2 = \ell$. 
$\Pi_2(z;\ell)$ is small and can be power-expanded in $1/\ell$.  
\opeqn 
 \Pi_2 (z;\ell)= - \left(\frac{1}{\ell} + \frac{\bar z }{\ell^2} \right)
        + {\cal O}(1/\ell^3)
\cleqn
Because of the vanishing dipole moment, the perturbing lens behaves as a point 
mass.
The DSTD lens equation (\ref{eqDSTD}) can be renormalized so that the 
unit distance is given by $r_{E1}$ and the
critical curve of the unperturbed lens 1 is given by $|z|=1$.
\opeqn 
 \omega = z - \int\dfrac{d^2 x^\prime \sigma_1(x^\prime)}
              {\bar z - \bar x^\prime - \tilde a_1 \overline{\Pi_2(z;\ell)}}
            - \frac{1}{\epsilon} \Pi_2(z;\ell)
\cleqn
where $x_2 =\ell$. 
This can be power-expanded with $\Pi_2$
as the small quantity assuming that the mass distribution of galaxy 1 is
well confined inside the critical curve.
\opeqn
 \omega - \frac{1}{\epsilon\ell} 
  = z - \Pi_{11} 
      + \frac{1}{\epsilon\ell^2}\bar z
      + \left(\frac{\tilde a_1}{\ell} + \frac{\tilde a_1 z}{\ell^2}\right) \Pi_{12}
      - \frac{\tilde a_1^2}{\ell^2} \Pi_{13} 
\cleqn
where
\opeqn
 \Pi_{1n} \equiv \int \frac{d^2 x^\prime \sigma_1(x^\prime)}
                           {(\bar z - \bar x^\prime)^n}
\cleqn
$\Pi_{11}$ is the deflection due to the galaxy mass 1 and the third 
term is the constant shear due to the perturber. There are also two 
other terms, proportional to $\Pi_{12}$ and $\Pi_{13}$ respectively,
that depend on the double scattering parameter $\tilde a_1$. 
It would be an error to ignore these last two terms if the (last scattering) 
perturbing galaxy is at a different distance.

The caustic of the main galaxy is usually of a finite size unless its 
projected mass distribution is circularly symmetric. The perturbation
by another galaxy will change the shape and size of the caustic, and
it is difficult to handle it algebraically in general. We leave the 
perturbations of finite size caustics for future work.

\appendix

\section{The Binary Lens with $\ell >> 1$}

With $d=1$, $\epsilon = M_1/M_2$, and the binary lens equation 
is obtained in which the main lens is $M_2$.  
\opeqn
  \omega = z - \frac{\epsilon}{\bar z -\ell} - \frac{1}{\bar z}
   = z + \frac{\epsilon}{\ell} + \frac{\epsilon}{\ell^2} \bar z 
       - \frac{1}{\bar z} + {\cal O}(\ell^{-3})
\cleqn
where the lens positions are $x_2 =0$ and $x_1 = \ell > 0$. 
The resulting lens is made of a point mass ($\propto \bar z^{-1}$) 
and a constant shear ($\propto \bar z$); 
the source is shifted by $\epsilon/\ell$.
The Jacobian components are
\opeqn
f = 1 ; \qquad
g = \frac{\epsilon}{\ell^2} + \frac{1}{\bar z^2}\,,
\cleqn
and its determinant $J$ is 
\opeqn
 J = 1 - \frac{2\epsilon \cos 2\theta}{\ell^2 r^2} 
       - \frac{1}{r^4} + {\cal O}(\ell^{-3})
\label{eqBiJ}
\cleqn
where $z = r e^{i\theta}$. The critical condition is given by
\opeqn
  \frac{2\epsilon \cos 2\theta}{\ell^2} = r^2 - \frac{1}{r^2} . 
\cleqn
From the simple graph of the RHS,
 it can be seen that there is a solution space
in the neighborhood of $r=1$. Set $r = 1 + \delta$ and find 
$\delta$ using eq.(\ref{eqBiJ})  to obtain
\opeqn
  r = 1 + \frac{\epsilon \cos 2\theta}{2\ell^2} \,. 
\cleqn 
It is a peanut-shape curve squeezed along the imaginary axis
even though the small coefficient $\epsilon/2\ell^2$ makes
it difficult to discern from a circle. 
$\theta =0$, $\pi/2$, $\pi$, and $3\pi/2$ are the precusps.
It can be confirmed by calculating $\partial_- J$ as it was done
for an arbitrary $d$ in the main text.
The cusps are two on the real axis and two on the imaginaray axis.
In order to estimate the size of the caustic, measure the 
cusp-to-cusp distances on the real axis and on the imaginary axis.
\opeqn
  \Delta \omega_{real} = \omega_0 - \omega_{\pi}
   = \frac{4\epsilon}{\ell^2} 
\cleqn 
\opeqn
  \Delta \omega_{imag} = \omega_{\pi/2} - \omega_{3\pi/2}
   = -i \frac{4\epsilon}{\ell^2}
\cleqn
The quaroid is equilateral and its orientation is opposite to the
critical curve. 
The diagonal length of the quadroid will be comapred to
that  of the large separation DSTP lens caustics.
\opeqn
  |\Delta \omega_{real}| 
= |\Delta \omega_{imag}| = \frac{4\epsilon}{\ell^2}
 = \frac{4}{\ell^2}\frac{M_1}{M_2}
\cleqn

\begin{figure}
\plotone{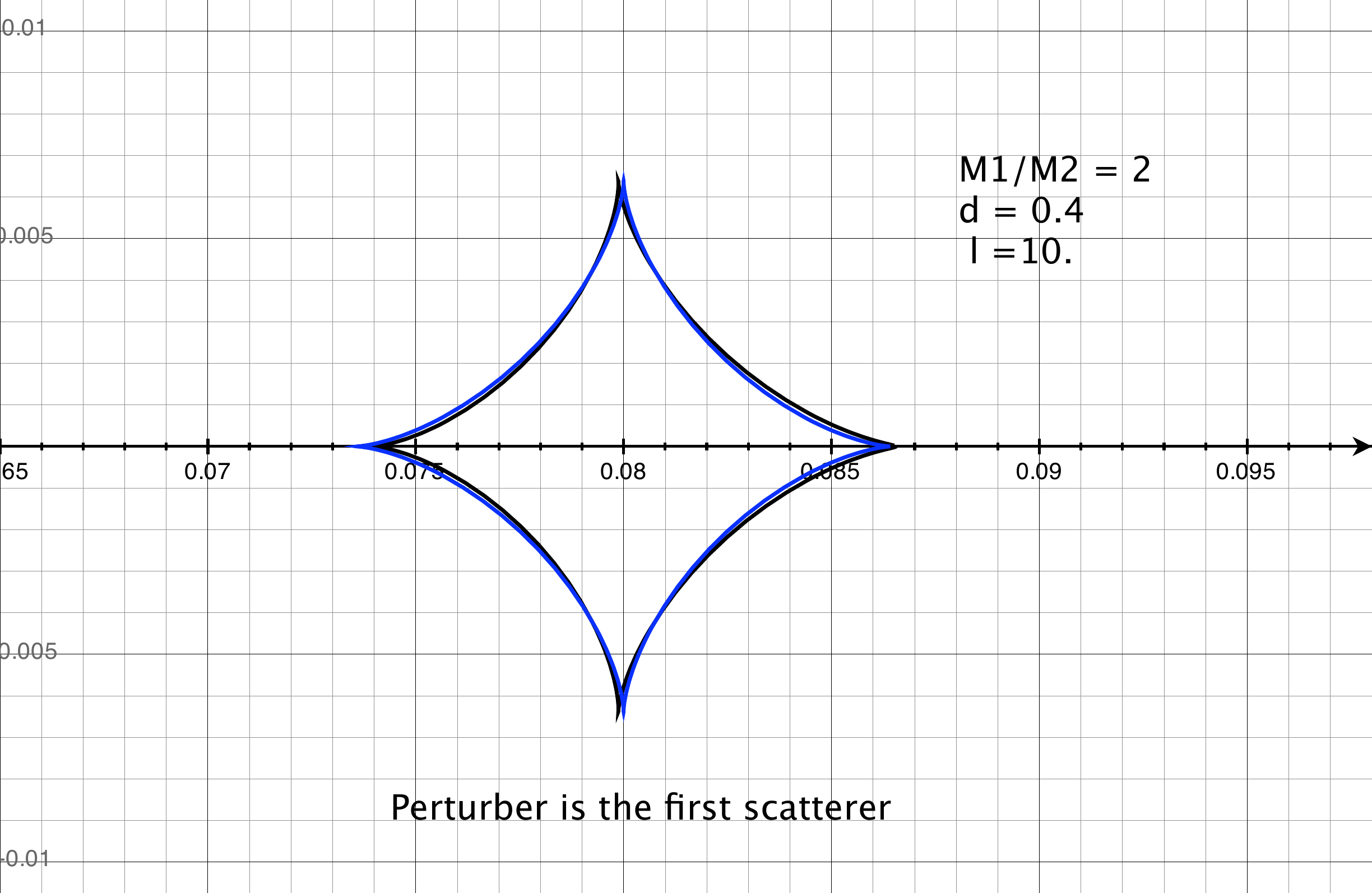}
\caption{
$M_2$ is the main lens at the origin:
The ``caustic curve" in blue is obtained by mapping 
the approximate critical curve using the approximate lens equation
with the perturber as the first scatterer. The perturber is to the
right on the real axis. 
The black ``caustic curve" is from the same approximate critical curve 
mapped by the exact DSTP lens equation. 
The exact caustic curve of the DSTP lens (not shown) is hardly
distinguishable from the black curve.
The size of the caustic is $\approx 0.013$ in agreement with the 
analytic formula.
}
\label{fig_cfirst}
\end{figure}

\begin{figure}
\plotone{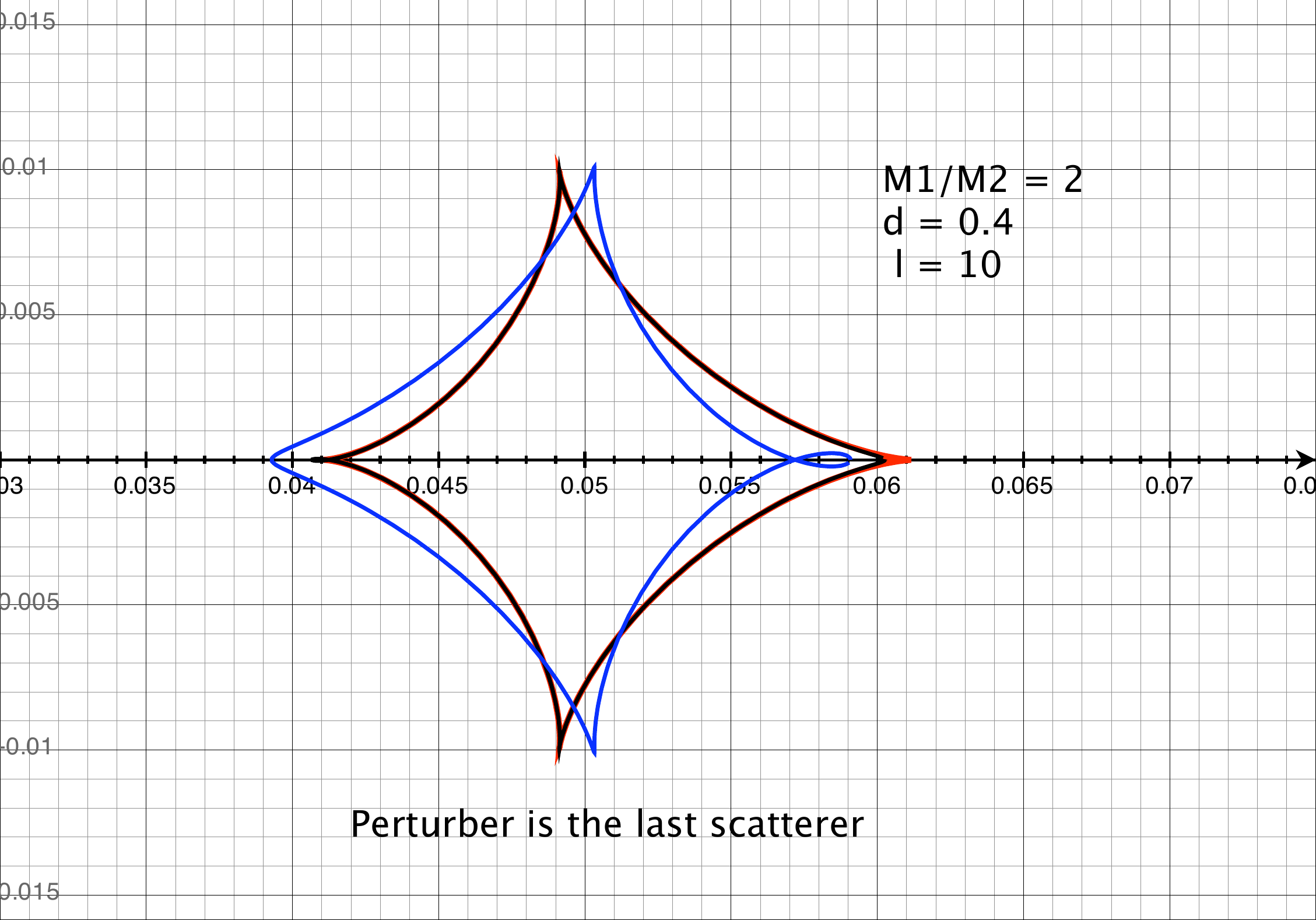}
\caption{
$M_1$ is the main lens at the origin:
The ``caustic curve" in blue is obtained by mapping the approximate
critical curve using the approximate lens equation with the perturber
as the last scatterer. The perturber is to the right on the real axis.
The black ``caustic curve" is from the same 
approximate critical curve mapped by the exact DSTP lens equation.
The exact caustic caustic curve of the DSTP lens is shown in thick 
red. The slight difference of the black curve from the red curve can 
be discerned in the area around the real axis toward the perturber.  
The ``caustic" sizes are $\approx 0.02$ in agreement with the analytic
formula.  
}
\label{fig_clast_j12}
\end{figure}

\begin{figure}
\plotone{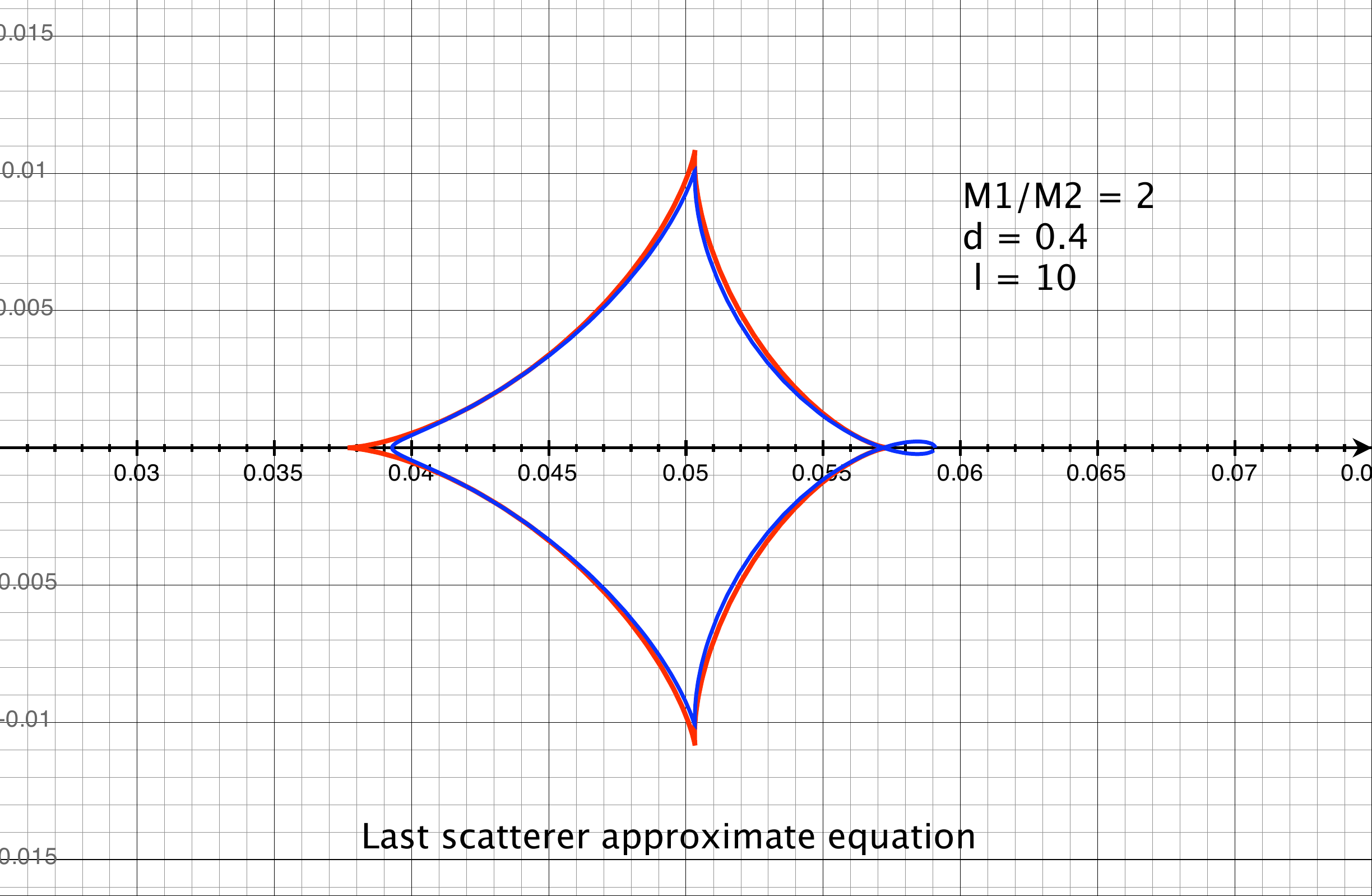}
\caption{
The ``caustic curve" in blue is the same as in fig.\ref{fig_clast_j12}.
The exact caustic curve of the approximate lens equation for the last
scattering perturber is shown in red. 
}
\label{fig_clast_j8}
\end{figure}

\end{document}